\documentclass[aps,prb,superscriptaddress,footinbib,onecolumn,preprint]{revtex4} 
\usepackage{graphicx}
\usepackage{dsfont}
\usepackage[latin1]{inputenc}
\usepackage{version} 
\usepackage{color}
\usepackage{amsmath}
\usepackage{braket}
\usepackage{natbib}
\usepackage{upgreek}
\usepackage{graphicx}
\usepackage{dcolumn}
\usepackage{bm}
\usepackage{bigints}
\usepackage{xcolor}  
\usepackage[pagebackref=true]{hyperref} 
\hypersetup{
    bookmarks=true,         
    unicode=false,          
    pdftoolbar=true,        
    pdfmenubar=true,        
    pdffitwindow=false,     
    pdfstartview={FitH},    
    pdftitle={Protecting a Spin Ensemble against Decoherence in the Strong-Coupling Regime of Cavity QED},    
    pdfauthor={Stefan Putz},     
    pdfkeywords={keyword1} {key2} {key3}, 
    pdfnewwindow=true,      
    colorlinks=true,       
    linkcolor=red,          
    citecolor=blue,        
    filecolor=magenta,      
    urlcolor=cyan,           
}

\makeindex
\begin{document}
\title{Protecting a Spin Ensemble against Decoherence in the Strong-Coupling Regime of Cavity QED} 
\author{S. Putz$^\S$}
\email[]{sputz@ati.ac.at}
\affiliation{Vienna Center for Quantum Science and Technology, Atominstitut, Vienna University of Technology, Stadionallee 2, 1020 Vienna, Austria}
\affiliation{Zentrum f\"ur Mikro- und Nanostrukturen, Vienna University of Technology, Floragasse 7, 1040 Vienna, Austria}
\author{D.~O. Krimer$^\S$}
\email[]{dmitry.krimer@gmail.com}
\affiliation{Institute for Theoretical Physics, Vienna University of Technology, Wiedner Hauptstrasse 8-10/136, 1040 Vienna, Austria}
\author{R. Ams\"uss}
\affiliation{Vienna Center for Quantum Science and Technology, Atominstitut, Vienna University of Technology, Stadionallee 2, 1020 Vienna, Austria}
\author{A. Valookaran}   
\affiliation{Vienna Center for Quantum Science and Technology, Atominstitut, Vienna University of Technology, Stadionallee 2, 1020 Vienna, Austria}
\author{T. N\"obauer}   
\affiliation{Vienna Center for Quantum Science and Technology, Atominstitut, Vienna University of Technology, Stadionallee 2, 1020 Vienna, Austria}
\author{J. Schmiedmayer}
\affiliation{Vienna Center for Quantum Science and Technology, Atominstitut, Vienna University of Technology, Stadionallee 2, 1020 Vienna, Austria}
\author{S. Rotter}   
\affiliation{Institute for Theoretical Physics, Vienna University of Technology, Wiedner Hauptstrasse 8-10/136, 1040 Vienna, Austria}
\author{J. Majer}
\affiliation{Vienna Center for Quantum Science and Technology, Atominstitut, Vienna University of Technology, Stadionallee 2, 1020 Vienna, Austria}
\affiliation{Zentrum f\"ur Mikro- und Nanostrukturen, Vienna University of Technology, Floragasse 7, 1040 Vienna, Austria}
\let\thefootnote\relax\footnotetext{$^\S$These authors contributed equally to this work (S.P. experiment, D.O.K. theory)}
\date{\today}
\label{par:abstract}
\begin{abstract} 
\textbf{Hybrid quantum systems based on spin ensembles coupled to superconducting microwave cavities are promising candidates for robust experiments in cavity quantum electrodynamics (QED) and for future technologies employing quantum mechanical effects.\cite{Zhu2011,Ze2013,Kubo2011,Amsuess2011} Currently the main source of decoherence in these systems is inhomogeneous spin broadening, which limits their performance for the coherent transfer and storage of quantum information.\cite{Kurucz2011,Sandner2012,Diniz2011} Here we study the dynamics of a superconducting cavity strongly coupled to an ensemble of nitrogen-vacancy centers in diamond. We experimentally observe for the first time, how decoherence induced by a non-Lorentzian spin distribution can be suppressed in the strong-coupling regime -- a phenomenon known as ``cavity protection''\cite{Kurucz2011,Diniz2011}. To demonstrate the potential of this effect for coherent control schemes, we show how appropriately chosen microwave pulses can increase the amplitude of coherent oscillations between cavity and spin ensemble by two orders of magnitude.}
\end{abstract}
\maketitle
\label{par:intro}

The processing of quantum information requires special devices that can store and manipulate quantum bits. Hybrid quantum systems\citep{Ze2013} combine the advantages of different systems in order to overcome their individual physical limitations. In this context superconducting microwave cavities have emerged as ideal tools for realizing strong coupling to qubits \citep{Wallraff2004,Majer2007,Kubo2010,Kubo2011,Schuster2010,Amsuess2011,Probst2013} allowing to store and retrieve excitations on the single photon level\citep{Kubo2012,Saito2013}. For the storage of quantum information the negatively charged Nitrogen-Vacancy (NV) centers in diamond show great potential, especially due to their long coherence times (up to one second \cite{Bar-Gill2013}) and due to the combination of microwave and optical transitions which makes them an easily accessible and controllable qubit\cite{Childress2006}. Coherently passing quantum information between such a spin and a cavity requires that they are strongly coupled to each other. As has recently been shown\cite{Kubo2010,Amsuess2011,Schuster2010,Probst2013}, this limit can be reached by collective coupling to a large spin ensemble, in which case the coupling strength is increased by the square root of the ensemble size. This collective coupling comes with a considerable downside though: in a solid state environment a spin is always prone to inhomogeneous broadening. In particular for an ensemble of NV centers magnetic dipolar interaction with excess nuclear and electron spins in the diamond crystal leads to an inhomogeneous broadening of the spin transition \cite{Stanwix2010}, which acts as the dominant source of decoherence. Overcoming this limitation is a considerable challenge for which several theoretical proposals have been put forward recently\cite{Kurucz2011,Diniz2011}. Implementations of these concepts rely on the specific shape of the inhomogeneous spectral spin distribution $\rho(\omega)$ of the NV center ensemble. Here we demonstrate, based on a non-Lorentzian spectral spin distribution the predicted but yet unobserved ``cavity protection effect'' \cite{Kurucz2011,Diniz2011} in an explicitly time-dependent study.

\label{par:experiment}

\begin{figure}[!ht]
\centering
\includegraphics[width=0.7\textwidth]{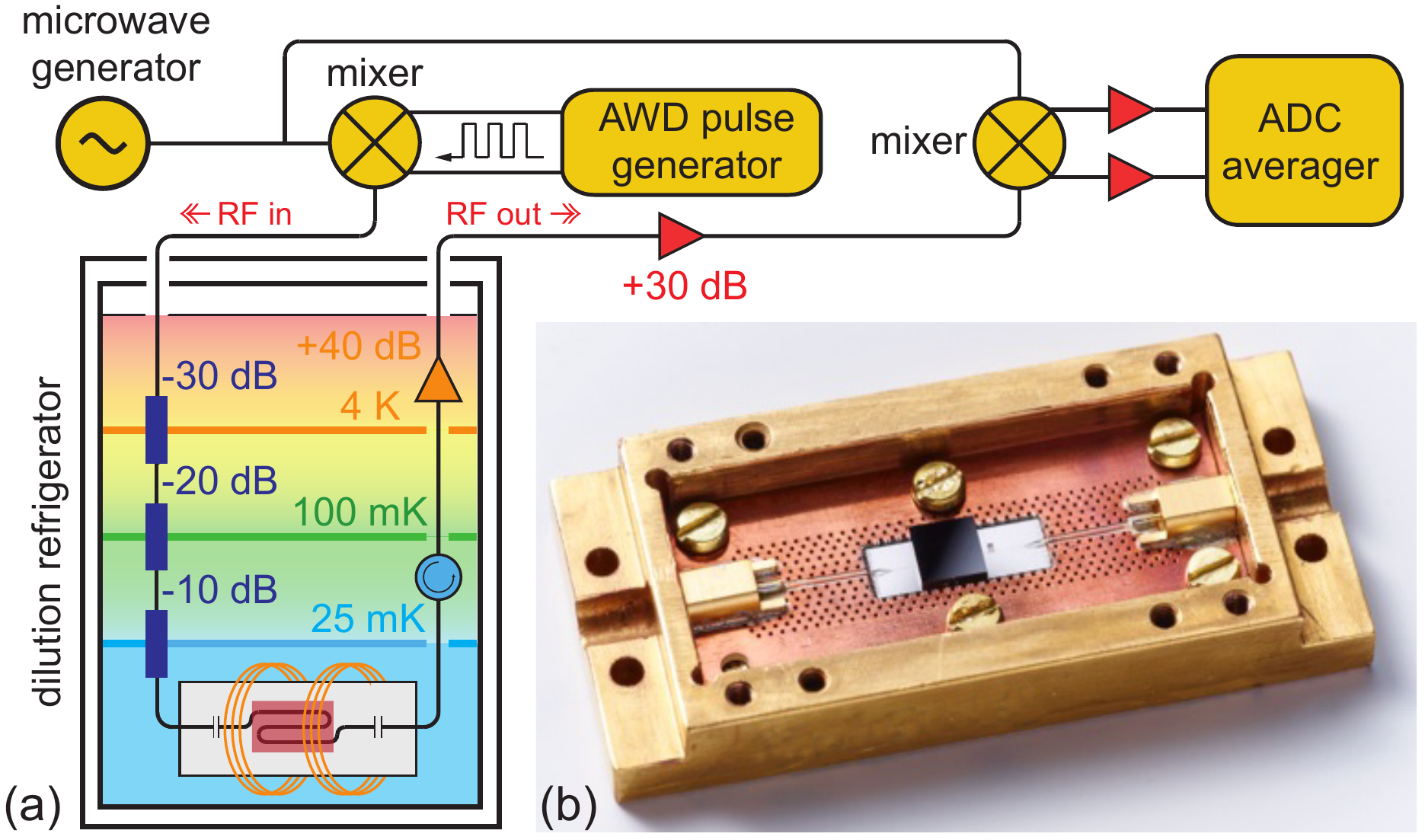}
\caption{\textbf{Experimental setup.} \textbf{(a)} The superconducting coplanar waveguide resonator with the diamond on top is cooled to $\sim 25$ mK in a dilution refrigerator. In our homodyne detection measurements, the input microwave signal is split into two paths, both serving as a reference signal as well as for testing and controlling our experiment. Outside the cryostat both signal paths are combined by a frequency mixer and the quadratures I and Q are recorded with a fast analog-to-digital converter with sub-nanosecond time resolution. \textbf{(b)} Photograph of a superconducting microwave cavity with an enhanced neutron irradiated type Ib synthetic diamond (black) on top, encased by a gold plated copper sample box.}
\label{fig:figure1}
\end{figure}

Our experiment is performed in a standard dilution refrigerator with the corresponding setup being sketched in Fig \ref{fig:figure1}a and a picture of the resonator with a synthetic diamond on top shown in Fig \ref{fig:figure1}b. To avoid thermal excitations we cool the entire setup to a temperature of 25 mK, where the estimated thermal spin polarization is of the order of 99\%. By applying an external magnetic field $|\mathbf{B}|=9.4$ mT through a set of two superconducting Helmholtz coils we Zeeman-tune the NV spin ensemble into resonance with the cavity. Our resonator has a fundamental resonance at $\omega_c/2\pi=2.6899$ GHz with a quality factor of $Q=3060$. To excite and probe the coupled system we inject microwave pulses into the cavity and perform time-resolved transmission spectroscopy by a fast homodyne detection setup with sub-nanosecond time resolution. The number of microwave photons in the cavity remains at or below $\sim 10^{6}$, which is very low compared to the number of $\sim 10^{12}$ NV spins involved in the coupling, ensuring that the Holstein-Primakoff\cite{Primakoff1939} approximation is valid for describing our experiments.

Our starting point to account for the dynamics of a single-mode cavity coupled to a spin ensemble is the Tavis-Cummings Hamiltonian\cite{Tavis68}, which reads in the rotating wave approximation
\begin{eqnarray}
H=\hbar\omega_ca^{\dagger}a+\frac{\hbar}{2}\sum_j^N\omega_j\sigma_j^z+\text{i}\hbar\sum_j^N\left[g_j\sigma_j^-a^{\dagger}-g_j^*\sigma_j^+a\right]-\text{i}\hbar\left[\eta(t) a^{\dagger}\text{e}^{-\text{i}\omega_p t}-\eta(t)^* a\text{e}^{\text{i}\omega_p t}\right]\,.
\label{Hamilt_fun}
\end{eqnarray}
The first and second term stand for the uncoupled resonator with frequency $\omega_c$ and for the spin ensemble with frequencies $\omega_j$, centered around $\omega_s$, respectively. The third and the last term describe the cavity spin interaction with coupling strength $g_j$ as well as the driving electromagnetic field injected into the cavity with amplitude $\eta(t)$ and frequency $\omega_p$. The collective coupling to a large number of spins allows us to enter the strong-coupling regime of QED, for which the interaction term is commonly reduced to a collective term\cite{Emary2003} $\Omega(S^-a^{\dagger}-S^+a)$, where the collective spin operators read $S^\pm=\frac{1}{\sqrt{N}}\sum_j^N\sigma_j^\pm$. The prefactor $\Omega^2=\sum_j^Ng_j^2$ stands for an effective coupling strength, which scales up a single cavity spin interaction typically on the order of $g_j\sim 2\pi \cdot 12$ Hz by a factor $\sqrt{N}$\cite{Verdu2009,Kubo2011,Amsuess2011}. In this formulation the effective spin-waves that are excited by the cavity mode can be identified as bright collective Dicke states which are effectively damped by the coupling to subradiant states in the ensemble\cite{Kubo2012,Dicke54}. To accurately describe the corresponding dynamics we also need to take into account the full non-Lorentzian spectral spin distribution\cite{Diniz2011} $\rho(\omega)=\sum_j g_j^2 \delta(\omega-\omega_j)/\Omega^2$. We achieve this by setting up a Volterra integral equation (see appendix), $A(t)=\int\limits_0^t d\tau \int  d\omega\, {\cal K}(\rho(\omega); t-\tau)  A(\tau)+{\cal F}(t)$, for the cavity amplitude $A(t)=\langle a(t)\rangle$. This includes a memory kernel  ${\cal K}(t-\tau)$ responsible for the non-Markovian feedback of the NV ensemble on the cavity and the function ${\cal F}(t)$ which describes the contribution from an external drive and initial spin excitation. In the following, the cavity amplitude $|A(t)|^2$, calculated with this approach for stationary and pulsed driving fields, will be compared to its experimental counterpart, i.e., the time-resolved microwave intensity measured in transmission through the cavity.
\label{par:results}

\begin{figure}[!ht]
\centering
\includegraphics[]{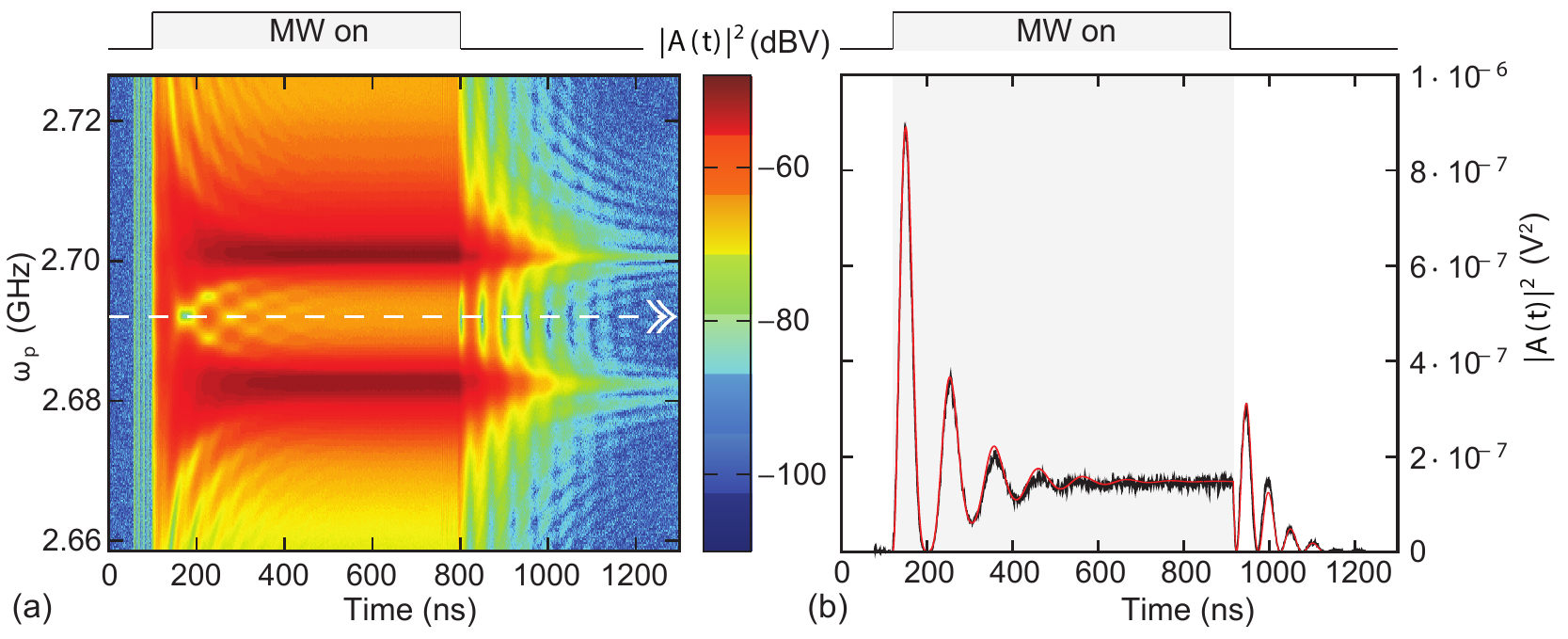}
\caption{\textbf{Time domain measurements of the cavity transmission.} \textbf{(a)} Transmission of a rectangular microwave pulse through the cavity vs.~time and probe frequency $\omega_p$ (the spins are on resonance with the cavity, $\omega_s=\omega_c$). The observation of a strong mode-splitting into the two hybridized modes $\ket{\Psi_\pm}\approx\frac{1}{\sqrt{2}}(\ket{0}_c\ket{1}_s\pm\ket{1}_c\ket{0}_s)$ (see dark red enhancements split by $\Omega_R=2 \pi \cdot 19.2$ MHz) confirm that the system is in the deep strong-coupling regime. \textbf{(b)} The dynamics at the resonant probe frequency $\omega_p=\omega_s=\omega_c$ [white dashed line in (a)] is compared with the theoretical prediction for the cavity probability amplitude $|A(t)|^2$ (experiment: black, theory: red). Excellent agreement is achieved when incorporating the correct non-Lorentzian spectral spin-distribution. After switching on the pulse, the system exhibits damped Rabi oscillations with frequency $\Omega_R$ that equilibrate at a stationary state. After switching off the pulse, the cavity amplitude first decays from the stationary state and then features a pronounced overshoot corresponding to a strongly non-Markovian release of the energy stored in the spin ensemble back into the cavity.}
\label{fig:figure2}
\end{figure}

First, to demonstrate that our experiment is in the strong-coupling regime (having $\omega_s=\omega_c$) we apply a rectangular microwave pulse sufficiently long ($800$~ns~$\gg 2\pi/\Omega$) to drive the system into a steady-state with varying probe frequency $\omega_p$. Fig \ref{fig:figure2}a shows that two effective eigenstates of the coupled system emerge in the transmission,\linebreak$\ket{\Psi_\pm}\approx\frac{1}{\sqrt{2}}(\ket{0}_c\ket{1}_s\pm\ket{1}_c\ket{0}_s)$, corresponding to the symmetric and antisymmetric superposition of the cavity and spin eigenstates, respectively. Strong coupling is secured since the Rabi splitting between these states $\Omega_R=2\pi\cdot 19.2 $ MHz is much larger than the total decay rate of the system $\Gamma=2\pi \cdot 3.0$ MHz (FWHM). The latter consists of a cavity decay rate, $\kappa=2\pi \cdot 0.8$ MHz (FWHM), as well as of a spin decay rate which contains a negligibly small spin dissipation $\gamma\rightarrow 0$ and a dominant contribution from the inhomogeneous broadening of the spin ensemble. Detailed spectroscopic measurements of the stationary transmission \cite{Sandner2012} reveal that the spectral function $\rho(\omega)$ which accurately captures the broadening is neither Lorentzian nor Gaussian, but has the intermediate form of a q-Gaussian\citep{Sandner2012} (see appendix). As shown in Fig \ref{fig:figure2}b our explicitly time dependent theoretical description yields excellent quantitative agreement with the experimental data, using such a q-Gaussian distribution function with a linewidth of $\gamma_q=2\pi\cdot9.4$ MHz (FWHM), a shape parameter $q=1.39$ and an effective coupling strength $2\cdot\Omega=2\pi\cdot17.2$ MHz. After turning on and switching off the microwave pulse coherent Rabi oscillations occur between the cavity and the spin ensemble, which we reproduce accurately including their damping. Interestingly, the first Rabi peak shows a pronounced overshoot after switching off the microwave drive, at which the energy stored in the spin ensemble is coherently released back into the cavity. These oscillations are a hallmark of the non-Markovian character of the system dynamics in the strong-coupling regime for which an accurate knowledge of the memory-kernel ${\cal K}(t-\tau)$ in our Volterra equation is essential.

\begin{figure}[!ht]
\centering
\includegraphics[angle=0,width=0.8\columnwidth]{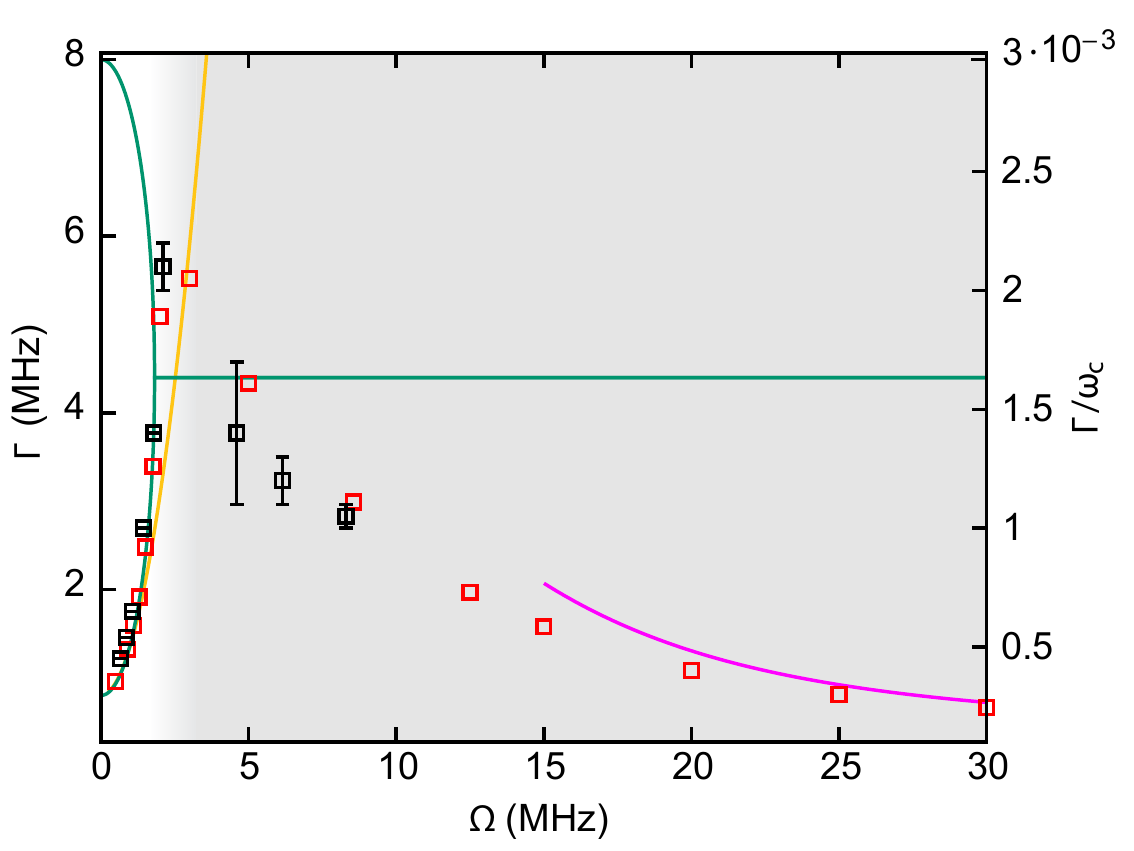}
\caption{\textbf{Characterization of the decay dynamics as a function of coupling strength.} For weak coupling, the decay rate $\Gamma$ of the cavity probability amplitude, $|A(t)|^2$, increases as a function of the coupling strength $\Omega$. For strong coupling, this trend reverses, showing a protection of the system against decoherence. {\it Black symbols}: experimentally observed decay rates. {\it Red symbols}: decay rates extracted from the full numerical calculations. {\it Orange curve}:  decay rate, $\Gamma=2 [\kappa+\pi \Omega^2 \rho(\omega_s)]$, derived under the Markovian approximation.  {\it Green curve}: characteristic decay rates,  $\Gamma_{1,2}=[-2(\Delta+\kappa)\pm\sqrt{(2\Delta-\kappa)^2-16 \Omega^2}]/4$ under the assumption of a Lorentzian distribution of the spin density. {\it Magenta curve}: analytical estimate of $\Gamma$ in the ultrastrong-coupling regime. The background color indicates at which coupling strength $\Omega$ the system undergoes a transition from the Markovian (white) to the non-Markovian (gray) regime.}
\label{fig:figure3}
\end{figure}

A first signature of the non-Lorentzian line shape of our spectral spin distribution $\rho(\omega)$ is that the period of the Rabi oscillations ($T_R=2\pi/\Omega_R=52$ ns) in Fig \ref{fig:figure2}b is not equal to the inverse effective coupling strength $\pi/\Omega=58$ ns. In other words, our hybrid cavity-spin system cannot be modeled as two coupled damped harmonic oscillators as in the case of a purely Lorentzian spin distribution. Especially for spectral distributions $\rho(\omega)$ which fall off faster than $1/\omega^2$ in their tails, an increasing coupling strength reduces the decay rate $\Gamma$ and protects the system against decoherence - hence the name ``cavity protection effect''\citep{Kurucz2011,Diniz2011}. Since the tails of our q-Gaussian spin distribution satisfy this required fast decay, we now have the possibility to probe this exceptional behavior in the experiment for the first time. We measure the decay rate $\Gamma$ of the cavity amplitude from a steady state (Fig~\ref{fig:figure2}b) for different coupling strengths $\Omega$. As we show in Fig~\ref{fig:figure3}, the values of $\Gamma(\Omega)$ vary over almost one order of magnitude in a strongly non-monotonic fashion: In the weak-coupling regime the decay rate $\Gamma$ increases with growing coupling strength $\Omega$ due to the Purcell-effect\cite{Purcell1946} as the cavity mode increasingly couples to the spin ensemble. Entering the strong-coupling regime, this trend reverses and $\Gamma$ decreases with growing $\Omega$. To highlight this remarkable phenomenon, we plot in Fig~\ref{fig:figure3} also the behavior for a Lorentzian spin distribution, for which $\Gamma(\Omega)$ is constant in the strong-coupling limit. Performing a Laplace transform of our Volterra equation we find  that in the limit of very strong coupling ($\Omega\to\infty$) the decay rate takes the following closed analytical form $\Gamma=\kappa+\pi \Omega^2 \rho(\omega_s\pm \Omega)$ (in agreement with a stationary analysis\cite{Diniz2011}). While the maximally reachable value of $\Omega=2\pi \cdot 8.6$ MHz in our device already leads to a considerable reduction of $\Gamma$ by 50$\%$ below its maximum, our numerical results (see Fig~\ref{fig:figure3}) predict a further reduction of the decay rate with increasing coupling strength by an order of magnitude.

\begin{figure}[!ht]
\centering
\includegraphics[]{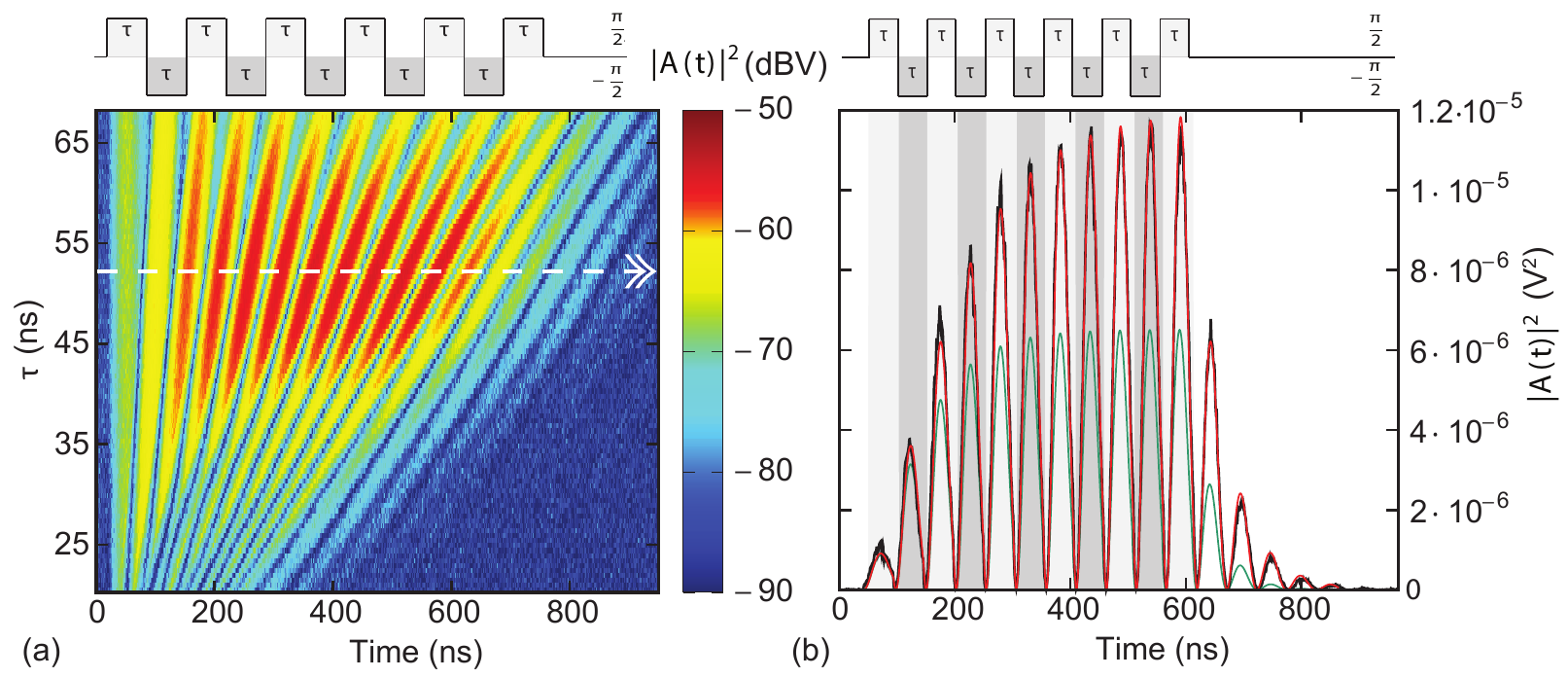}
\caption{\textbf{Enhancement of the cavity transmission intensity by pulsed driving.} \textbf{(a)} The cavity transmission of eleven successive rectangular microwave pulses with carrier frequency $\omega_p=\omega_c=\omega_s$, phase-switched by $\pi$, as a function of time and pulse duration $\tau$ (see top panel for the pulse shape). \textbf{(b)} Dynamics at the largest enhancement of the cavity transmission corresponding to a pulse duration of $\tau=52$~ns equal to the Rabi period $T_R=2\pi/\Omega_R$. After switching off the probe signal the system settles back to the ground state through damped Rabi oscillations. Excellent agreement between experiment (black curve) and theory (red curve) is found, using the same system parameters as in Fig.~\ref{fig:figure3}. A Lorentzian spin distribution in the theoretical calculations (green curve) leads to a considerably smaller enhancement due to the absence of the cavity-protection effect.}
\label{fig:figure4}
\end{figure}

In a next step, we demonstrate that the ``cavity-protection effect'' can also be employed for the realization of coherent-control schemes. In particular, we address a central question when dealing with coherently driven spin ensembles, which is how to achieve high excitation levels in the spin ensemble with limited driving powers\citep{Grezes2014,Sigillito2014}. In a simplified picture of two coupled harmonic oscillators this can be achieved by a drive modulated with the inverse of the effective coupling strength. To realize this for the non-Lorentzian spectral spin distribution of our ensemble a pulsed driving is required to match the Rabi frequency $\Omega_R$ rather than the effective coupling strength $2\Omega$, which quantities are quite different from each other. We thus probe our setup by a driving field with a carrier frequency $\omega_p=\omega_c=\omega_s$ and a periodical modulation with tunable period $\tau$. Realizing the latter with a simple periodic sign-change of the carrier signal, we find that this driving scheme produces giant oscillations in the transmission (see Fig~\ref{fig:figure4}a) corresponding to a coherent exchange of energy between the cavity and the spin ensemble. A maximum oscillation amplitude occurs exactly at the point where the modulation period $\tau$ coincides with the inverse of the Rabi splitting  $2\pi/\Omega_R$. Note that at this resonant driving the steady-state oscillation amplitude in the transmission signal (see Fig~\ref{fig:figure4}b) exceeds the stationary amplitude (see Fig~\ref{fig:figure2}b) by two orders of magnitude, although the net power applied to the cavity is exactly the same in both cases. Our approach demonstrates how to sustain coherent oscillations and how to reach considerably high excitation amplitudes of the spin ensemble without using strong driving powers. For comparison, we also plot in Fig~\ref{fig:figure4}b both the results for a q-Gaussian as well as for a Lorentzian spin-density, which clearly shows the substantially lower excitation amplitudes for the Lorentzian case. This clear signature of the ``cavity-protection effect'' paves the way for the realization of sophisticated coherent-control schemes in the strong-coupling regime of QED.

\label{par:conclusion}
In conclusion, we present the first experimental demonstration of the so-called ``cavity-protection effect'', which shields an inhomogeneously broadened spin ensemble strongly coupled to a cavity mode against its own decoherence. As we demonstrate in our time dependent study, this effect substantially reduces the decay rates in our hybrid quantum system and can further be improved by increasing the collective coupling strength $\Omega$. To highlight the potential of this effect for the implementation of coherent-control schemes, we reveal how an appropriately chosen pulse sequence can excite and maintain giant coherent oscillations between the cavity and the spin ensemble.

\section*{Appendix: Volterra equation for the cavity amplitude}
\normalsize
We start from the Hamiltonian (1) of the main article and derive the Heisenberg operator equations (limit of zero temperature), for the cavity and spin operators, $\dot a=i [{\cal H},a]-\kappa a$, $\dot \sigma_k^-=i [{\cal H},\sigma_k^-]-\gamma \sigma_k^-$, respectively. Here $\kappa$ and $\gamma$ stand for the total cavity and spin losses, respectively. We then write a set of equations for the expectation values in the frame rotating with the probe frequency $\omega_p$, using the commonly used Holstein-Primakoff-approximation, $\langle \sigma_k^z \rangle \approx -1$, which is valid if the number of the excited spins is small compared to the ensemble size (which is the case for all experimental results reported in the main article). Denoting $A(t)\equiv \langle a(t)\rangle$ and $B_k(t)\equiv\langle\sigma_k^-(t)\rangle$, we end up with the following set of first-order ODEs with respect to the cavity and spin amplitudes
\begin{subequations}
\begin{eqnarray}
\label{Eq_a_Volt}
\dot{A}(t) & = & -\left[\kappa-i(\omega_c-\omega_p)\right]A(t) + \sum_k
g_k  B_k(t)-\eta(t), \\
\label{Eq_bk_Volt}
\dot{B}_k(t) & = & -\left[\gamma+i(\omega_k-\omega_p)\right] B_k(t) - g_k A(t).
\end{eqnarray}
\end{subequations}
Note, that the size of our spin ensemble is very large (typically $N\sim 10^{12}$) and individual spins are distributed around a certain mean frequency $\omega_s$. We can thus go to the continuum limit by introducing the continuous spectral density as $\rho(\omega)=\sum_k g_k^2 \delta(\omega-\omega_k)/\Omega^2$ (see, e.g. \onlinecite{Diniz2011}), where $\Omega$ is the collective coupling strength of the spin ensemble to the cavity and $\int d\omega\rho(\omega)=1$. In what follows we will replace any discrete function $F(\omega_k)$ by its continuous counterpart, $F(\omega)$: $F(\omega_k) \rightarrow \Omega^2 \int d\omega \rho(\omega) F(\omega)$. By integrating Eq.~(\ref{Eq_bk_Volt}) in time, each individual spin amplitude, $B_k(t)$, can formally be expressed in terms of the cavity amplitude, $A(t)$. By plugging the resulting equation into Eq.~(\ref{Eq_a_Volt}) and assuming that initially all spins are in the ground state, $B_k(t=0)=0$, we arrive at the following integro-differential Volterra equation for the cavity amplitude ($\omega_c=\omega_s$)
\begin{eqnarray}
\dot A(t)=-\kappa A(t)-\Omega^2 \int d\omega \rho(\omega) \int\limits_{0}^t d\tau
e^{-i(\omega-\omega_c-i\gamma)(t-\tau)}A(\tau)-\eta(t), 
\label{Eq_rigor}
\end{eqnarray}
Note that in the $\omega_p$-rotating frame the rapid oscillations presented in the original Hamiltonian (1) are absent, so that the time variation of $\eta(t)$ in Eq.~(\ref{Eq_rigor}) is much slower as compared to $1/\omega_p$. 

For a proper description of the resulting dynamics, it is essential to capture the form of the spectral density $\rho(\omega)$ realized in the experiment as accurately as possible. Following \onlinecite{Sandner2012}, we take the $q$-Gaussian function for that purpose 
\begin{eqnarray}
\label{rho_w_Eq}
\rho(\omega)=C\cdot\left[1-(1-q)\dfrac{(\omega-\omega_s)^2}{\Delta^2}\right]^{
\dfrac{1}{1-q}},
\end{eqnarray}
characterized by the dimensionless shape parameter $1<q<3$ which yields the form of a Lorentzian and Gaussian distribution, for $q=2$ and for $q\rightarrow 1$, respectively.  Here $C$ is a normalization constant which is easily obtained numerically; the full-width at half-maximum (FWHM) of $\rho(\omega)$ is given by $\gamma_q=2\Delta\sqrt{\dfrac{2^q-2}{2q-2}}$.

After integrating Eq.~(\ref{Eq_rigor}) in time, performing some algebraic manipulations and assuming that the cavity is initially empty, $A(t=0)=0$, we derive the following equation for the cavity amplitude
\begin{eqnarray}
\label{Volt_eq}
A(t)=\int\limits_0^t d\tau {\cal K}(t-\tau) A(\tau)+{\cal F}(t),
\end{eqnarray}
which contains the kernel function ${\cal K}(t-\tau)$,
\begin{eqnarray}
\label{Volt_eq_K}
{\cal K}(t-\tau)=\Omega^2 \bigintsss\!\!\! d\omega\,
\dfrac{\rho(\omega) \left[e^{-i (\omega-\omega_c-i (\gamma-\kappa))(t-\tau)}-1\right]
}{i (\omega-\omega_c-i (\gamma-\kappa))}\cdot e^{-\kappa(t-\tau)},
\end{eqnarray}
and the function ${\cal F}(t)$,
\begin{eqnarray}
\label{Volt_eq_F}
{\cal F}(t)=\int\limits_0^t d\tau\, \eta(\tau)\cdot e^{-\kappa(t-\tau)}.
\end{eqnarray}
Despite its seemingly simple form, Eq.~(\ref{Volt_eq}) is not trivial to solve in practice, even numerically. The reasons are twofold: First, the result of the integration for $A(t)$ at time $t$ depends on the amplitude $A(\tau)$ calculated at all earlier times, $\tau<t$ (memory effect). Second, the kernel function ${\cal K}(t-\tau)$ contains the integration with respect to frequency, which is costly in terms of computational time. (Note that such an integration has to be performed for each $t$ and $\tau<t$.) The smallest possible time scale in our problem is given by $T=2\pi/\omega_p\sim 0.4\,$ns. To achieve a very good accuracy of the calculations for the results presented in Figs.~2,4 from the main article, we solve the equation on a mesh with uniform spacing, choosing a time step $dt \sim 0.05\,$ns (see e.g. \onlinecite{NumRec} for more details about the method). The direct discretization of ${\cal K}(t-\tau)$ on the time interval of the order of $\upmu$s (typical time of measurements) leads to a high-dimensional matrix (of a size typically exceeding $10^4 \times 10^4$), which, together with the integration with respect to frequency, makes the problem computationally intractable by way of a direct numerical solution. To overcome this problem and to speed up the calculations drastically, we divide the whole time integration into many successive subintervals, $T_{ n}\leq t \leq T_{n+1}$, with $n=1,2,...$. Such a time division might, in principle, be implemented arbitrarily but we choose it to be adapted to our experimental realization. Specifically, the driving amplitude is unchanged within each subinterval, so that in our case it is given by
\begin{equation} 
\eta_n=
\begin{cases}
\eta & n=1,3,5,...\\ -\eta & n=2,4,6,...
\end{cases}.
\end{equation}

In this way the result of integration at the $n$-th time interval, $A^{(n)}(T_{n+1})$, enters as an initial condition for the integration during the $(n+1)$-th time interval, $A^{(n+1)}(T_{n+1})$. Finally, we end up with the following recurrence relation (time runs within $T_{ n}\leq t \leq T_{n+1}$ for $n=1,2,3,...$)
\begin{eqnarray}
\label{Volt_eq_itter}
A^{(n)}(t)=\int\limits_{T_n}^t d\tau {\cal K}(t-\tau) A^{(n)}(\tau)+{\cal F}^{(n)}(t),
\end{eqnarray}
where the kernel function ${\cal K}(t-\tau)$ is defined by Eq.~(\ref{Volt_eq_K}) and
\begin{eqnarray}
\nonumber
{\cal F}^{(n)}(t)&=&A^{(n-1)}(T_n)e^{-\kappa(t-T_n)}+\Omega^2 e^{-\kappa(t-T_n)}
\bigintsss\!\!\! d\omega\,
\dfrac{\rho(\omega) \left[e^{-i (\omega-\omega_c-i  (\gamma-\kappa))(t-T_n)}-1\right]
}{i (\omega-\omega_c-i (\gamma-\kappa))}\cdot {\cal I}_n(\omega)+
\\
&&\dfrac{{\cal\eta}_n}{\kappa}\cdot\left[1-e^{-\kappa(t-T_n)} \right]
\end{eqnarray}
Remarkably, the memory about previous events enters both through the amplitude $A^{(n-1)}(T_n)$ and through the function  
\begin{eqnarray}
{\cal I}_n(\omega)=e^{-i(\omega-\omega_p-i\gamma)(T_n-T_{n-1})}{\cal I}_{n-1}(\omega)+
\int\limits_{T_{n-1}}^{T_n} d\tau e^{-i(\omega-\omega_p-i\gamma)(T_n-\tau)}A^{(n-1)}(\tau).
\end{eqnarray}
In accordance with the initial conditions introduced above ($t=T_1=0$), $A(T_1)=0$ and $ {\cal I}_1(\omega)=0$.

The above technique allows us to solve Eq.~(\ref{Volt_eq}) accurately while being very efficient in terms of computational time. We have tested the accuracy of our numerical results by varying the discretization both in time and frequency in a wide range obtaining excellent agreement with the experimental results shown in Figs.~2,4 of the main paper and thereby confirming the accuracy of our method.

\textbf{~\\Acknowledgements }
\label{par:ack}
We would like to thank C. Koller, F. Mintert, P. Rabl, H. Ritsch, K. Sandner, and M. Trupke for helpful discussions. 
The experimental effort has been supported by the TOP grant of TU Vienna. S.P. acknowledges support by the Austrian Science Fund (FWF) through Project No.~W1243 (Solids4Fun), D.K. and S.R. through Projects No. F25-P14 (SFB IR-ON), No. F49-P10 (SFB NextLite).

\textbf{~\\Author contributions }
\label{par:contrib}
S.P., R.A., A.V., T.N., J.S, and J.M. designed and set up  the experiment. S.P., R.A., and A.V. carried out the measurements with supervision by J.M.. D.O.K. and S.R. devised the theoretical framework and performed the calculations. S.P., D.O.K., S.R. and J.M. wrote the first manuscript draft to which all authors suggested improvements.

\label{biblio}
\bibliographystyle{plain}
\newpage
\bibliography{mybib}
\clearpage

\end{document}